\documentclass[12pt]{article}
\usepackage{graphicx,multirow}
\usepackage{latexsym}
\usepackage{amsmath,amsfonts,amssymb}

\begin{document}
\begin{center}
{\bf CASIMIR FORCE BETWEEN DIELECTRIC MEDIA WITH FREE CHARGES }

\vspace{1cm} Johan S. H{\o}ye\footnote{johan.hoye@ntnu.no}

\bigskip
Department of Physics, Norwegian University of Science and
Technology, N-7491 Trondheim, Norway

\bigskip
Iver Brevik\footnote{iver.h.brevik@ntnu.no}

\bigskip

Department of Energy and Process Engineering, Norwegian University
of Science and Technology, N-7491 Trondheim, Norway

\vspace{1cm} \today
\end{center}

\begin{abstract}
The statistical mechanical approach to Casimir problems for
dielectrics separated by a vacuum gap turns out to be compact and
effective. A central ingredient of this method is  the effect of
interacting fluctuating dipole moments of the polarizable
particles. At arbitrary temperature the path integral formulation
of quantized particles, developed by H{\o}ye-Stell and others, is
needed. At high temperature - the limit considered in the present
paper - the classical theory is however sufficient. Our present
theory is related to an idea put forward earlier by Jancovici and
{\v S}amaj (2004), namely to evaluate the Casimir force between
parallel plates invoking an electronic plasma model and the
Debye-H{\"u}ckel theory for electrolytes. Their result was
recently recovered by H{\o}ye (2008), using a related statistical
mechanical method. In the present paper we generalize this by
including a constant permittivity in the description. The present
paper generalizes our earlier theory for parallel plates (1998),
as well as for spherical dielectrics (2001). We also consider the
Casimir force between a polarizable particle and a conductor with
a small density of charges, finding agreement with the result
recently derived by Pitaevskii (2008).

\end{abstract}


\section{Introduction}

The  large interest in the Casimir effect in recent years has
resulted in a flood of papers, most of which are concerned with
the field theoretical approach to the problem. This is quite
natural, and is in accordance with the spirit in the original
paper of Casimir paper \cite{casimir48}. Some general reviews of
the Casimir effect can be found in
Refs.~\cite{bordag01,milton04,lamoreaux05,buhmann07}. Quantum
field theoretical methods have proved to be quite effective even
in the presence of dielectric media, although there are
fundamental problems here, especially as regards perfect conductor
boundary conditions at sharp surfaces \cite{birrell82}.

Now there exists an alternative and probably less known
alternative route to derive a theory for the Casimir effect,
namely to start from statistical mechanics and regard the effect
to be due to interacting fluctuating dipole moments of polarizable
particles. As shown by Brevik and H{\o}ye, the Casimir force
between a pair of polarizable point particles can be recovered
\cite{brevik88}. In this context, the path integral formulation of
quantized particle systems was utilized. The method had  earlier
been applied to polarizable fluid systems, by H{\o}ye, Stell, and
others \cite{hoye81}. Subsequent generalizations led to the well
known Lifshitz formula for parallel plates \cite{hoye98}. Similar
evaluations were performed in Refs.~\cite{hoye01,hoye03}.

We wish to emphasize, as a general remark, that the quantum
statistical mechanical method is quite compact and effective. For
instance, the Casimir free energy due to the dispersion force
between two polarizable particles as given by Eq. (5.15) in
\cite{brevik88} is derived in a way which according to our view
 is simpler than the conventional field theoretical method. The
latter involves  fourth order approximation of perturbation theory
\cite{berestetskii82}. Moreover, we ought to point out that the
statistical mechanical approach opens new perspectives in the
sense that one avoids quantizing the electromagnetic field itself.
Instead, the field is playing the role of an agent permitting the
interaction between  polarizable particles. The two pictures are
physically equivalent \cite{brevik88,hoye98,hoye01,hoye03}.

The topic with which we will be concerned with below, is related
to earlier work of Jancovici and {\v S}amaj \cite{jancovici04}.
They realized that it should be possible to evaluate the Casimir
force between metallic plates by employing an electronic plasma
model. They considered accordingly charged particles at low
density in a neutralized background, limiting themselves to the
classical (i.e., the high temperature limit) for which the
Debye-H{\"u}ckel theory of electrolytes is fully applicable.
Calculating the pair correlation function, and from that the local
ionic density at the surfaces of the plates, they were able to
recover the conventional Casimir result for real metals in the
high temperature limit. The issue of temperature corrections to
the Casimir force has been subject to a lively discussion in the
contemporary literature \cite{sernelius00}. The ionic plasma has
also been extended to the quantum mechanical case by use of the
path integral formalism from a statistical mechanical viewpoint
\cite{buenzli05}.

The recent paper of H{\o}ye \cite{hoye08} reconsidered the ionic
plasma in the classical limit, using the  statistical mechanical
method in a different way to obtain the Casimir force. The
correlation function was used to directly evaluate the average
force between pairs of particles in the two plates and then the
total force was found by integration. This is the same method as
used in Refs.~\cite{brevik88,hoye98}. The result was found to be
in agreement with Ref.~\cite{jancovici04}. A notable feature of
this approach is that it demonstrates how the modification of the
density profile at the surface is a perturbing effect that can be
neglected to leading order.

In Ref.~\cite{hoye08} the two parallel plates (or slabs) were
assumed to be filled with ions, but no dielectric properties due
to the medium itself were envisaged. The present paper presents an
extension and generalization of the theory in \cite{hoye08}, in
the sense that the dielectric properties are included. That means,
we introduce  permittivities $\varepsilon$ in the slabs. The two
permittivities are taken to be constant, and equal. Magnetic
properties in the media are omitted. In the gap region between the
plates, we assume a vacuum. Our method consists in replacing
polarizations by equivalent charges. That is, the polarization
$\bf P$ is taken equivalent to an induced charge density
\begin{equation}
\rho_I=-\bf \nabla\cdot P. \label{1}
\end{equation}
This follows from the connection between the  electric field $\bf
E$, the polarization $\bf P$, and the charge density $\rho_c$,
\begin{equation}
{\bf \nabla \cdot E}+4\pi {\bf \nabla \cdot P}=4\pi \rho_c.
\label{2}
\end{equation}
When evaluating the Casimir force based upon the ionic fluid
correlation function, the induced charge can be considered as a
free charge. Accordingly, no separate charged dipole and
dipole-dipole correlation functions are  needed. In addition, the
dielectric properties are expressed via the influence of the
permittivity upon the charge-charge correlation function.

In the next section we establish the basic formalism for the pair
correlation function and thereafter, in section 3, derive the
Casimir force as a generalization of the expression found earlier
for the case where charges were absent \cite{hoye98}. In section 4
we consider the force between a polarizable particle and a slab
(half-space) containing free charges. The treatment in that
section is related to recent work of Pitaevskii
\cite{pitaevskii08}. Finally, in section 5 we consider the Casimir
interaction between spherical dielectric shells and extend the
theory given earlier in Ref.~\cite{hoye01} to the case where there
are free charges present.

\section{The correlation function}
 To obtain the correlation function
we start from the Ornstein-Zernike (OZ) equation \cite{hoye08}
\begin{equation}
h({\bf r}_2, {\bf r}_1)=c({\bf r}_2, {\bf r}_1)+\int c({\bf r}_2,
{\bf r}')\rho({\bf r}')h({\bf r}', {\bf r}_1) d{\bf r}', \label{3}
\end{equation}
here extended to non-homogeneous fluids.  In this equation $h({\bf
r}_2, {\bf r}_1)$ is the correlation function, $c({\bf r}_2, {\bf
r}_1)$ is the direct correlation function, and $\rho({\bf r}')$ is
the number density of charges.

It may be appropriate to give a brief account of  the background
for this equation. The
 correlation function or pair correlation function
$h({\bf r}_2, {\bf r}_1)$ is related to the pair distribution
function $g({\bf r}_2, {\bf r}_1)$ through
\[ \rho({\bf r}_1)\rho({\bf r}_2)h({\bf r}_2, {\bf r}_1)=g({\bf
r}_2,{\bf r}_1)-\rho({\bf r}_1)\rho({\bf r}_2). \] The $g({\bf
r}_2, {\bf r}_1)$ is the probability density for one particle to
occupy the position ${\bf r}_2$ while another particle occupies
the position ${\bf r}_1$. For an ideal gas with uncorrelated
particles, $g=\rho({\bf r}_1)\rho({\bf r}_2)$. Thus the $h$
expresses deviations from the ideal gas value. For a uniform fluid
$\rho({\bf r})=\rho= $constant, and $h\rightarrow h({\bf r}_2-{\bf
r}_1)$. The above equation as originally introduced by Ornstein
and Zernike \cite{ornstein14}  serves as a definition of the
direct correlation function $c({\bf r}_2, {\bf r}_1)$. In their
investigation they noted that the direct correlation function $c$
for a fluid was closely related to the  pair interaction
 itself. In view of this, Eq.~(\ref{3}) can be
given a simple interpretation: The resulting pair correlation
function $h$ is the result of a direct correlation $c$ plus
correlations via other particles as expressed by the integral.

 For weak long range forces, the   $c({\bf r}, {\bf r}')$ is
to leading order related to the interaction $\psi$ in a simple way
\cite{hemmer64},
\begin{equation}
c({\bf r}, {\bf r}')=-\beta \psi({\bf r}, {\bf r}'), \label{4}
\end{equation}
where $\beta=1/k_B T$, $T$ is the temperature, and $k_B$ is
Boltzmann's constant. This follows from the $\gamma$-ordering
studied in Ref.~\cite{hemmer64}, where $\gamma$ is the inverse
range of interaction and the  limit $\gamma \rightarrow 0$ is
considered. In the present case this becomes exact for large
separations $\bf r-r'$ with particles located in separate plates.
However, for particles at close separation (in the same plate)
there will be deviations. For low densities they can be neglected,
in accordance with the Debye-H{\"u}ckel theory for electrolytes.
For higher densities such deviations will mainly change the
inverse Debye-shielding length. This contributes only to a minor
change of the effective separation between the plates
\cite{hoye08}. We find no reason to consider this further here,
especially since we consider semiconductors with low density of
charges.

As the  interaction  follows from the electrostatic potential
between two charges, we have
\begin{equation}
{\bf \nabla}^2 c({\bf r}, {\bf r}')=4\pi \beta
q_c^2\frac{1}{\varepsilon({\bf r}')}\delta({\bf r}-{\bf r}'),
\label{5}
\end{equation}
with $\varepsilon ({\bf r}')$ the permittivity at position $\bf
r'$, and $q_c$ the ionic charge assuming one component for
simplicity. It is then assumed that the ions are neutralized by  a
uniform background of counter ions.

Equation (\ref{3}) can now be rewritten
\[ \nabla^2 \Phi-4\pi \beta
q_c^2\frac{1}{\varepsilon({\bf r})}\rho({\bf r})\Phi=-4\pi
\frac{1}{\varepsilon ({\bf r}_0)}\delta({\bf r}-{\bf r}_0), \]
\begin{equation}
h({\bf r}, {\bf r}_0)=-\beta q_c^2\Phi, \label{6}
\end{equation}
where $\Phi$ is the electrostatic potential. Here we  have
replaced ${\bf r}_2$ and ${\bf r}_1$ by $\bf r$ and ${\bf r}_0$,
respectively. With parallel plates the number density is
\begin{equation}
\rho ({\bf r})=\left\{\begin{array}{lll} \rho, & z<0 \\
                                0,  & 0<z<a \\
                                \rho, & a<z
                                \end{array}
                                \right. \label{7}
                                \end{equation}

\noindent with equal densities ($\rho$ constant) in the two media.
By Fourier transform in the $x$ and $y$ directions, Eq.~(\ref{6})
becomes
\begin{equation}
\left(\frac{\partial^2}{\partial
z^2}-k_\perp^2-\kappa_\varepsilon^2\right)\hat{\Phi}=-4\pi
\frac{1}{\varepsilon (z_0)}\delta(z-z_0), \label{8}
\end{equation}
where the hat denotes Fourier transform. Further, we have
introduced the quantity $\kappa_\varepsilon$, defined by
\begin{equation}
\kappa_\varepsilon^2=4\pi \beta q_c^2\rho/\varepsilon \label{9}
\end{equation}
in the two media and $\kappa_\varepsilon=0$ in the vacuum gap
$0<z<a$. Physically, $\kappa_\varepsilon$ is the inverse
Debye-H{\"u}ckel shielding length. The symbol ${\bf k}_\perp$ is
the wave vector transverse to the $z$ direction. The solution of
Eq.~(\ref{8}) can be written in the form
\begin{equation}
\hat{\Phi}=2\pi e^{q_\kappa z_0}\left\{\begin{array}{lll}
\frac{1}{\varepsilon q_\kappa}e^{-q_\kappa z}+Be^{q_\kappa z}, &
z_0<z<0 \\
Ce^{-qz}+C_1e^{qz}, & 0<z<a \\
De^{-q_\kappa z}, & a<z
\end{array}
\right. \label{10}
\end{equation}
where $q=k_\perp,~
q_\kappa=\sqrt{k_\perp^2+\kappa_\varepsilon^2}$. [ For $z<z_0$,
the $-q_\kappa(z-z_0)$ is to be replaced by $-q_\kappa|z-z_0|$ in
the first line of Eq.~(\ref{10}).] As the boundary conditions
require that $\hat{\Phi}$ and $\varepsilon
\partial {\hat \Phi}/\partial z$ are continuous, one finds for the
coefficient of main interest
\begin{equation}
D=\frac{4qe^{(q_\kappa -q)a}}{(\varepsilon q_\kappa
+q)^2(1-Ae^{-2qa})},
 \quad A=\left(\frac{\varepsilon
q_\kappa-q}{\varepsilon q_\kappa+q}\right)^2. \label{11}
\end{equation}
With this the pair correlation function for free charges, for
$z_0<0$ and $z>a$, is
\begin{equation}
{\hat h}(k_\perp, z, z_0)=-2\pi \beta q_c^2De^{-q_\kappa(z-z_0)}.
\label{12}
\end{equation}

\section{Casimir force}

To obtain the Casimir force the ionic interaction $\psi=\psi(r)$
in vacuum is needed. This is the Coulomb potential $\psi=q_c^2/r$.
Its full Fourier transform is ${\hat \psi}=4\pi q_c^2/k^2$
consistent with Eq.~(\ref{5}) with $\varepsilon=1$. With
$k^2=k_\perp^2+k_z^2$ this can be transformed backwards to obtain
(recall that $q=k_\perp$)
\begin{equation}
{\hat \psi}(k_\perp, z-z_0)=2\pi q_c^2\frac{e^{-q|z-z_0|}}{q}.
\label{13}
\end{equation}
With $q^2=k_\perp^2=k_x^2+k_y^2$, $dk_xdk_y=2\pi qdq$, we now
obtain for the Casimir force per unit area, for the free ions,
\begin{equation}
 f=\frac{\rho^2}{2\pi}\int
\hat{h}(q,z,z_0)\psi_z'(q,z-z_0)qdqdzdz_0, \label{add3}
\end{equation}
where $\psi_z'=\partial \psi/\partial z$, the hat denoting Fourier
transform with respect to the $x$ and $y$ coordinates. (We have
used $\int fg dxdy=\int \hat{f}\hat{g}dk_x dk_y/(2\pi)^2$ and the
translational symmetry along the $xy$ plane.) Cf. Eq.~(6) in
Ref.~\cite{hoye08}. Equation (\ref{add3}) follows from from the
general structure of the expressions for the Casimir free energy
and force established in Refs.~\cite{brevik88} and \cite{hoye98}.
In those references the interaction was the dipole-dipole
interaction while the correlation function was the dipole-dipole
correlation function. In the present case, by contrast, those
quantities are replaced by the charge-charge interaction and the
charge-charge correlation function.

With $z-z_0 =u_1+u_2+a$, $\kappa^2=\varepsilon
\kappa_\varepsilon^2$ we  obtain
\[f  =\frac{\rho^2}{2\pi}\int_0^\infty (-2\pi \beta q_c^2)D(2\pi
q_c^2)\int_0^\infty \int_0^\infty
e^{-(q_\kappa+q)(u_1+u_2+a)}du_1du_2qdq \]
\begin{equation}
=-\frac{\kappa^4}{8\pi \beta}\int_0^\infty
\frac{De^{-(q_\kappa+q)a}}{(q_\kappa+q)^2}qdq. \label{14}
\end{equation}
However, this expression is missing the contribution to the force
due to fluctuating dipole moments. In the evaluation of the pair
correlation function we avoided this problem by use of the
permittivity. Here, we may use the free charge picture of dipole
moments where the induced charge density in terms of polarization
is given by the expression (\ref{1}). Alternatively, we can use
the method developed by H{\o}ye and Stell in their analysis of the
ion-dipole fluid mixture where all the correlations functions
between ions and dipoles were obtained for a uniform fluid
\cite{hoye78}.

The polarization of a particle at position $\bf r$ can be written
as ${\bf P}={\bf s}_i\delta ({\bf r}-{\bf r}_i)$, where ${\bf
s}_i$ is the dipolar moment. The induced charge $\rho_I$ as given
by Eq.~(\ref{1}) will replace the $q_c$ factors in both the
interaction (\ref{13}) and the correlation function (\ref{12}).
But this charge $\rho_I=-{\bf \nabla \cdot P}$ is distributed in
space and should be convoluted with the point particle interaction
$\hat{\psi}$ and the correlation function $\hat{h}$. Taking into
account the Fourier transform in the transverse directions and the
exponentials in the $z$ direction, together with the derivative of
the delta-function, we find that at each end of $\hat \psi$ and
$\hat h$ the $q_c$ should be replaced by
\[
 q_c\rightarrow {\bf h}\cdot {\bf s}_i, \quad {\bf h}=\{ik_x, ik_y,
\pm q\}\] and
\begin{equation}
 q_c\rightarrow {\bf h}_\kappa \cdot {\bf s}_i, \quad {\bf
h}_\kappa=\{ik_x, ik_y, \pm q_\kappa\}, \label{15}
\end{equation}
respectively. [For convenience we keep the notation of
Ref.~\cite{hoye98}. Thus this $\bf h$ should not be confused with
the correlation function $h$ introduced in Eq.~(\ref{3}).] Now one
needs thermal averages. There is no correlation between ionic
charges and dipole moments  ${\bf s}_i$ for the reference system
(i.e., with the ionic interaction absent). For polar or
polarizable particles we assume isotropy so that
\[
\langle {\hat s}_{ix}^2\rangle=\langle {\hat
 s}_{iy}^2\rangle=\langle { \hat s}_{iz}^2\rangle=\frac{1}{3},\quad
\langle { \hat s}_{ix}{ \hat s}_{iy}\rangle=0, {\rm etc.},
\]
where the hat denotes unit vectors. To obtain the Casimir force
 we need the average
\[ Q=\langle ({\bf h}\cdot {\hat {\bf s}}_i) ({\bf h}_\kappa \cdot {\hat {\bf s}}_i)^*\rangle =\frac{1}{3}(\bf h
\cdot {h}_\kappa^*) \]
\begin{equation}
=\frac{1}{3}(k_\perp^2+qq_\kappa)=\frac{1}{3}q(q_\kappa+q)
\label{16}
\end{equation}
($k_x^2+k_y^2= k_\perp^2=q^2$). Note that this is the same as the
square root of the result (6.27) of Ref.~\cite{hoye98}, with
$q_\varepsilon$ replaced by $q_\kappa$. Further, the dipolar
density $\rho_d$ replaces the ionic density $\rho$. We may now
define the quantity $y$ via
\begin{equation}
3y=\frac{4\pi}{3}\beta \rho_d \langle s_i^2\rangle=4\pi \rho_d
\alpha, \label{17}
\end{equation}
where $\alpha$ is the polarizability.  Thus for the dipoles we
altogether have the replacement
\begin{equation}
4\pi \beta \rho q_c^2\rightarrow 4\pi \beta \rho_d\langle
s_i^2\rangle Q=9yQ. \label{18}
\end{equation}
This should be modified, however, by another factor
$(\varepsilon-1)/3y$, which is present in Eq.~(5.5) in
Ref.~\cite{hoye98} or in Eq.~(33) in Ref.~\cite{hoye78}. This may
be regarded as a contribution from correlations between
neighboring dipole moments in a reference system where the ideal
dipole-dipole interaction (times $-\beta$) has been subtracted
from the direct correlation function. Thus we have
\begin{equation}
4\pi \beta \rho q_c^2 \rightarrow
3(\varepsilon-1)Q=(\varepsilon-1)q(q_\kappa+q). \label{19}
\end{equation}
From the free ions we have $4\pi \beta \rho
q_c^2=\kappa^2=\varepsilon \kappa_\varepsilon^2$. Adding the
contribution (\ref{19}) we obtain
\[ \kappa^2\rightarrow \varepsilon
\kappa_\varepsilon^2+(\varepsilon-1)q(q_\kappa+q) \]
\begin{equation}
=\varepsilon
(q_\kappa^2-q^2)+(\varepsilon-1)q(q_\kappa+q)=(\varepsilon
q_\kappa-q)(q_\kappa+q). \label{20}
\end{equation}
Altogether, the resulting Casimir force will be the one in which
the $\kappa^4$ in Eq.~(\ref{14}) is replaced with the square of
expression (\ref{20}). We find
\begin{equation}
f=-\frac{1}{2\pi \beta}\int_0^\infty
\frac{Ae^{-2qa}}{1-Ae^{-2qa}}q^2dq, \label{21}
\end{equation}
with $A$ given by Eq.~(\ref{11}). This expression is a simple
generalization of the situation with absence of dielectric
properties \cite{hoye08}, or absence of charges \cite{hoye98}.

\section{Force between a polarizable particle and a half-space
with charges}

This is a situation earlier considered by Pitaevskii
\cite{pitaevskii08}. We will reconsider it with the formalism
developed above. To simplify, the polarizable particle will be
considered to be confined in a thin plane or layer together with
other particles of the same kind at low density. Then the
electrostatic problem will be as before except for the boundary
condition where now one of the half-spaces (slabs) is removed and
replaced with the thin layer of polarizable particles of vanishing
density not influencing the electric field. Thus, instead of
Eq.~(\ref{10}), the solution of the electrostatic problem becomes
\begin{equation}
\hat{\Phi}=2\pi \left\{ \begin{array}{ll} \frac{1}{q}e^{-qz}+Be^{qz}, & 0<z<a \\
De^{-q_\kappa z}, & a<z
\end{array}
\right. \label{22}
\end{equation}
with the polarizable particle(s) located at $z_0=0$. From the
usual boundary conditions one finds
\begin{equation}
D=\frac{2e^{(q_\kappa-q)a}}{\varepsilon q_\kappa+q}. \label{23}
\end{equation}
The Casimir force is obtained by a modification of Eq.~(\ref{14}).
First, one of the integrations is reduced to a thin layer of width
$\Delta d$ at $u_1=0$. Secondly, the modification (\ref{20}) is
needed with dielectric media. Thus
\begin{equation}
4\pi \beta \rho q_c^2=\kappa^2\rightarrow (\varepsilon
q_\kappa-q)(q_\kappa+q). \label{24}
\end{equation}
So for the half-space
\begin{equation}
\kappa^2\rightarrow \kappa_2^2=(\varepsilon
q_\kappa-q)(q_\kappa+q), \label{25}
\end{equation}
while for the thin layer with no free charges,
$q_\kappa\rightarrow q$. Further, the permittivity $\varepsilon_1$
of the thin layer is directly related to the polarizability
$\alpha$ and the density of dipolar particles $\rho_1 (\rightarrow
0$) via
\begin{equation}
\varepsilon_1-1=4\pi \alpha \rho_1. \label{26}
\end{equation}
So for the thin layer
\begin{equation}
\kappa^2\rightarrow \kappa_1^2=2(\varepsilon_1-1)q^2=8\pi \alpha
\rho_1q^2. \label{27}
\end{equation}
Using Eqs.~(\ref{23})-(\ref{27}) in Eq.~(\ref{14}) we find (with
$\Delta d=\int du_1$)
\[ f=-\frac{\kappa_1^2\kappa_2^2\Delta d}{8\pi \beta}\int_0^\infty
\frac{De^{-(q_\kappa+q)a}}{q_\kappa+q}qdq \]
\begin{equation}
=-(\rho_1\Delta d)\frac{2\alpha}{\beta}\int_0^\infty
\frac{\varepsilon q_\kappa-q}{\varepsilon
q_\kappa+q}e^{-2qa}q^3dq. \label{28}
\end{equation}
Now the number of particles per unit area is $\rho_1\Delta d$, so
the force upon each of them is $f/(\rho_1\Delta d)$. The
corresponding interaction potential $V$ is determined from
$\partial V/\partial a=-f/(\rho_1\Delta d)$. Integration of
Eq.~(\ref{28}) thus gives
\begin{equation}
V=-\frac{\alpha}{\beta}\int_0^\infty \frac{\varepsilon
q_\kappa-q}{\varepsilon q_\kappa+q}\, e^{-2qa}\,q^2dq, \label{29}
\end{equation}
which is the result obtained by Pitaevskii \cite{pitaevskii08}.
See also the comment of Geyer {\it et al.}, and the reply of
Pitaevskii \cite{geyer09}.

\section{Free energy of concentric spherical bodies}

Consider an inner sphere (ball) with radius $a$ and an outer
sphere with inner radius $b$ $(>a)$ and outer radius at infinity.
The spheres have permittivity $\varepsilon$. Between the radii $a$
and $b$ there is a vacuum gap. So far this is the situation
considered by H{\o}ye {\it et al.}  \cite{hoye01}. in that paper
general expressions were found, for arbitrary temperature, for
nonmagnetic spheres including both the transverse magnetic and the
transverse electric modes for nonzero Matsubara frequencies. In
the present work we want to extend this to the situation where
also free charges are present. As in the parallel-plates situation
we have to restrict the evaluation to the electrostatic or the
classical high temperature case. With this limitation the
resulting free energy will be a straightforward extension of
previous results.

Employing spherical coordinates the solution for the potential
$\Phi$, Eq.~(\ref{6}), can be written
\begin{equation}
\Phi=\Phi_l(r)Y_{lm}(\theta, \varphi), \label{30}
\end{equation}
where $Y_{lm}$ are the spherical harmonics. The radially dependent
term can be written as
\begin{equation}
\Phi_l(r)=\left\{ \begin{array}{lll} e_\varepsilon+Bs_\varepsilon,
& r<a
\\
Ce+C_1s, & a<r<b \\
De_\varepsilon, & b<r.
\end{array}
\right. \label{31}
\end{equation}
With no free charges (i.e., $\kappa=0$), Eq.~(\ref{6}) is the
Laplace equation and the functions $e_\varepsilon$ and
$s_\varepsilon$ simplify to
\begin{equation}
e_\varepsilon \propto e \propto \frac{1}{r^{l+1}} \quad {\rm and}
\quad s_\varepsilon \propto s \propto r^l. \label{32}
\end{equation}
However, with free charges present in the media the
$\kappa_\varepsilon$ given above in Eq.~(\ref{9}) will enter
Eq.~(\ref{6}). Then the solution for the radial part becomes
Riccati-Bessel functions with imaginary argument. As $\Phi_l(r)$
should be finite at the origin and zero at infinity we can write
\begin{equation}
s_\varepsilon=rj_l(kr) \quad {\rm and} \quad
e_\varepsilon=rh_l^{(1)}(kr), \label{33}
\end{equation}
with
\begin{equation}
-k^2=\kappa_\varepsilon^2=4\pi \beta q_c^2\rho/\varepsilon
\label{34}
\end{equation}
in the media. In vacuum $k=i\kappa_\varepsilon \rightarrow 0$, and
we can put
\begin{equation}
e=\frac{1}{r^{l+1}}\quad {\rm and} \quad s=r^l \label{35}
\end{equation}
since proportionality factors are not needed here.

One notes that Eqs.~(\ref{31}) and (\ref{33}) are are precisely
those of Ref.~\cite{hoye01} for the situation with TM (transverse
magnetic) waves where Matsubara frequencies $K=-i\hbar \omega=2\pi
n/\beta$ ($n$ integer) were used. In that case the imaginary
values were   $k=\omega/c$ in vacuum and $k=\sqrt{\varepsilon}
\omega/c$ in the media. With this equivalence the free energy
associated with the mutual interaction between the inner and outer
spheres can be evaluated precisely as in Ref.~\cite{hoye01}. Since
the expression (\ref{31}) for $\Phi_l$ does not determine its
magnitude we can use the method in Secs.~IV and V of \cite{hoye01}
to obtain it indirectly, and thus obtain the eigenvalues of
interest.

For parallel plates we were able to do this directly, through
Eqs.~(\ref{15})-(\ref{20}) above, to obtain the force (\ref{21}).
By integration with respect to the separation $a$ the
corresponding free energy in that case is found to be
\begin{equation}
\beta F=\frac{1}{2}\frac{1}{(2\pi)^2}\int \ln (1-\lambda_q)2\pi
qdq, \label{36}
\end{equation}
 which identify the eigenvalues $\lambda_q=Ae^{-2qa}$.

 To obtain the corresponding eigenvalues $\lambda_{\varepsilon l}$
 with the concentric spheres we write, similarly as in
 \cite{hoye01},
 \begin{equation}
 D=\frac{D_0}{1-\lambda_{\varepsilon l}}. \label{37}
 \end{equation}
 The quantity $D$ is to be determined from the continuity of
 $\Phi_l$ and $\varepsilon \Phi_l'$ at the surfaces. This gives the
 equations
 \[ e_{a\varepsilon}+Bs_{a\varepsilon}=Ce_a+C_1s_a, \]
 \[ \varepsilon(e'_{a\varepsilon}+Bs'_{a\varepsilon})=Ce'_a+C_1s'_a,\]
 \[ Ce_b+C_1 s_b=De_{b\varepsilon}, \]
 \begin{equation}
 Ce'_b+C_1s'_b=\varepsilon De'_{b\varepsilon}, \label{38}
 \end{equation}
where subscripts $a$ and $b$ denote the radial positions and the
prime denotes differentiation with respect to $r$. [Due to a
calculational error in \cite{hoye01} the corresponding equations
(33) in that reference differ from those above in that
$\varepsilon$ is replaced by its inverse. The error may be related
to the somewhat involved discussion in \cite{hoye01} on the
conditions for the transverse magnetic field and the related
electric field for nonzero frequencies.] Solving for $D$ one also
needs $D_0$ to determine $\lambda_{\varepsilon l}$. As explained
in Sec. IV of \cite{hoye01}, this can be obtained by considering
single potential bonds between the two spheres. We then first
remove the inner sphere by putting $a=0$ to obtain from the two
last members of Eq.~(\ref{38})
\begin{equation}
D=D_0=c_2C, \quad {\rm with} \quad
c_2=\frac{e_b's_b-e_bs_b'}{\varepsilon
e_{b\varepsilon}'s_b-e_{b\varepsilon}s_b'}. \label{39}
\end{equation}
Then, removal of the outer sphere by putting $b=\infty$ and
$C_1=0$ yields from the two first members of Eq.~(\ref{38})
\begin{equation}
C=C_\infty =c_1, \quad {\rm with} \quad
c_1=\frac{\varepsilon(e_{a\varepsilon}s_{a\varepsilon}'-e_{a\varepsilon}'s_{a\varepsilon})}
{\varepsilon e_as_{a\varepsilon}'-e_a's_{a\varepsilon}},
\label{40}
\end{equation}
from which $D_0=c_1c_2$. Finally solving the full set of equations
(\ref{38}) one finds $D$. This further used in Eq.~(\ref{37})
yields
\begin{equation}
\lambda_{\varepsilon l}=\frac{(\varepsilon
s_as_{a\varepsilon}'-s_a's_{a\varepsilon})(\varepsilon
e_be_{b\varepsilon}'-e_b' e_{b\varepsilon})} {(\varepsilon
e_as_{a\varepsilon}'-e_a's_{a\varepsilon})(\varepsilon
e_{b\varepsilon}'s_b-e_{b\varepsilon}s_b')}. \label{41}
\end{equation}
With no free charges, $\kappa_\varepsilon=0$, by which
$e_\varepsilon=e=1/r^{l+1}$ and $s_\varepsilon =s= r^l$. Then
$\lambda_{\varepsilon l}$ simplifies to
\begin{equation}
\lambda_{\varepsilon
l}=\frac{(\varepsilon-1)^2(l+1)l}{(\varepsilon
l+l+1)[\varepsilon(l+1)+l]}\left( \frac{a}{b}\right)^{2l+1},
\label{42}
\end{equation}
as found in Ref.~\cite{hoye01}. The resulting free energy $F$ is
now the zero frequency  ($K=0$) TM mode term of equation (40) in
\cite{hoye01},
\begin{equation}
\beta F=\frac{1}{2}\sum_{l=1}^\infty (2l+1)\ln
(1-\lambda_{\varepsilon l}), \label{43}
\end{equation}
with expression (\ref{41}) inserted, where $k$ is given by
Eq.~(\ref{34}).

\section{Summary}
 With the use of a statistical mechanical method the Casimir force
 between dielectric slabs containing free ions (i.e.,
 semiconductors) has been evaluated in the high temperature
 classical limit.  Further, for high temperatures the  free
 energy of interaction between a polarizable particle and a
 semiconductor slab has been obtained.  Agreement with earlier
 results of Pitaevskii is found \cite{pitaevskii08}.
 Finally, in the same limit the free energy of interaction between
 two concentric semiconducting dielectric spheres separated by a
 vacuum gap is found.


\newpage
\begin{thebibliography}{99}

\bibitem{casimir48}
H. B. G. Casimir, Proc. K. Ned. Akad. Wet. {\bf 51}, 793 (1948).
\bibitem{bordag01}
M. Bordag, U. Mohideen and V. M. Mostepanenko, Phys. Rep. {\bf
353}, 1 (2001).
\bibitem{milton04}
K. A. Milton, J. Phys. A {\bf 37}, R209 (2004).
\bibitem{lamoreaux05}
S. K. Lamoreaux, Rep. Prog. Phys. {\bf 68}, 201 (2005).
\bibitem{buhmann07}
S. Y. Buhmann and D.-G. Welsch, Progr. Quantum Electron. {\bf 31},
51 (2007).
\bibitem{birrell82}
See, for instance, N. D. Birrell and P. C. W. Davies, {\it Quantum
Fields in Curved Space} (Cambridge University Press, Cambridge,
England, 1982); D. Deutsch and P. Candelas, Phys. Rev. D {\bf 20},
3063 (1979).
\bibitem{brevik88}
I. Brevik and J. S. H{\o}ye, Physica A {\bf 153}, 420 (1988).
\bibitem{hoye81}
J. S. H{\o}ye and G. Stell, J. Chem. Phys. {\bf 75}, 5133 (1981);
M. J. Thomson, K. Schweizer, and D. Chandler, J. Chem. Phys. {\bf
76}, 1128 (1982).
\bibitem{hoye98}
J. S. H{\o}ye and I. Brevik, Physica A {\bf 259}, 165 (1998).
\bibitem{hoye00}
J. S. H{\o}ye and I. Brevik, J. Stat. Phys. {\bf 100}, 223 (2000).
\bibitem{hoye01}
J. S. H{\o}ye, I. Brevik, and J. B. Aarseth, Phys. Rev. {\bf 63},
051101 (2001).
\bibitem{hoye03}
J. S. H{\o}ye, I. Brevik, J. B. Aarseth, and K. A. Milton, Phys.
Rev. E {\bf 67}, 056116 (2003).
\bibitem{berestetskii82}
See, for instance, V. B. Berestetskii, E. M. Lifshitz, and L. P.
Pitaevskii, {\it Quantum Electrodynamics} (Pergamon Press, Oxford,
1982), Sect. 85.
\bibitem{jancovici04}
B. Jancovici and L. {\v S}amaj, J. Stat. Mech. P08006 (2004).
\bibitem{sernelius00}
For some recent papers along these lines, see M. Bostr{\"o}m and
Bo E. Sernelius, Phys. Rev. Lett. {\bf 84}, 4757 (2000); I.
Brevik, S. A. Ellingsen, and K. A. Milton, New J. Phys. {\bf 8},
236 (2006); G. L. Klimchitskaya and V. M. Mostepanenko, Contemp.
Phys. {\bf 47}, 131 (2006); J. S. H{\o}ye, I. Brevik, S. A.
Ellingsen, and J. B. Aarseth, Phys. Rev. E {\bf 75}, 051127
(2007); G. L. Klimchitskaya and B. Geyer, J. Phys. A {\bf 41},
164032 (2008); G. L. Klimchitskaya, e-print arXiv:0902.4022
[cond-mat.other].
\bibitem{buenzli05}
P. R. Buenzli and Ph. A. Martin, Phys. Rev. E {\bf 77}, 011114
(2008).
\bibitem{hoye08}
J. S. H{\o}ye, in {\it The Casimir Effect and Cosmology}
(dedicated to the 70th anniversary of Prof. Iver H. Brevik, edited
by S. D. Odintsov {\it et al.}, Tomsk State Pedagogical
University, 2008), p. 117 [arXiv:0903.2975].
\bibitem{pitaevskii08}
L. P. Pitaevskii, Phys. Rev. Lett. {\bf 101}, 163202 (2008).
\bibitem{ornstein14}
L. S. Ornstein and F. Zernike, Proc. Acad. Sci. (Amsterdam) {\bf
17}, 793 (1914).
\bibitem{hemmer64}
P. C. Hemmer, J. Math. Phys. {\bf 5}, 75 (1964); J. L. Lebowitz,
G. Stell, and S. Baer, J. Math. Phys. {\bf 6}, 1282 (1965).
\bibitem{hoye78}
J. S. H{\o}ye and G. Stell, J. Chem. Phys. {\bf 68}, 4145 (1978).
\bibitem{geyer09}
B. Geyer, G. L. Klimchitskaya, U. Mohideen, and V. M.
Mostepanenko, preprint arXiv:0810.3243 v2 (to appear in Phys. Rev.
Lett.); L. P. Pitaevskii, preprint arXiv:0811.3081 v1.


\end{thebibliography}
\end{document}